\documentstyle[pre,aps,epsf,psfig,amssymb,floats,amssymb]{revtex}

\begin{document}
\flushbottom

\def\be{\begin{equation}}
\def\bea{\begin{eqnarray}}
\def\ee{\end{equation}}
\def\eea{\end{eqnarray}}
\def\ri{\rightarrow}
\def\ov{\overline}
\def\ra{\rangle}
\def\la{\langle}
\def\scr{\scriptscriptstyle}
\def\bottomfraction{0.5}

\twocolumn[\hsize\textwidth\columnwidth\hsize\csname @twocolumnfalse\endcsname

\title{
Scaling of waves in the Bak-Tang-Wiesenfeld sandpile model}
\author{D.\,V.~Ktitarev$^{1,2}$\footnote{On leave of absence from:
Laboratory of Computing Techniques,
JINR, Dubna,
141980 Russia},
        S.~L\"ubeck$^{2}$,
        P.~Grassberger$^{1}$ and
        V.\,B.~Priezzhev$^{3}$}

\address{
$^1$John von Neumann Institut f\"ur Computing,
Forschungszentrum J\"ulich, 52425 J\"ulich, Germany\\
$^2$Theoretische Physik,
Gerhard-Mercator-Universit\"at Duisburg,
47048 Duisburg, Germany \\
$^3$Laboratory of Theoretical Physics, Joint Institute for Nuclear Research,
Dubna, 141980 Russia}

\date{June 30, 1999}

\maketitle

\begin{abstract}
We study probability distributions of waves of topplings in the 
Bak-Tang-Wiesenfeld model on hypercubic lattices for dimensions $D\ge2$.
Waves represent relaxation processes which do not contain multiple toppling 
events. We investigate bulk and boundary waves by means of their correspondence
to spanning trees, and by extensive numerical simulations. While 
the scaling behavior of avalanches is complex and usually not governed by 
simple scaling laws, we show that the probability distributions
for waves display clear power law asymptotic behavior in perfect 
agreement with the analytical predictions. Critical exponents are obtained 
for the distributions of radius, area, and duration, of bulk and boundary waves. 
Relations between them and fractal dimensions of waves are derived. We confirm 
that the upper critical dimension $D_{\rm u}$ of the model is 4, and calculate 
logarithmic corrections to the scaling behavior of waves in $D=4$. In addition 
we present analytical estimates for bulk avalanches in dimensions $D\ge 4$ 
and simulation data for avalanches in $D\le 3$. For $D=2$ they seem not easy to 
interpret.
\end{abstract}


] 

\setcounter{page}{1}
\markright{\rm submitted to
Phys. Rev.~E  }
\thispagestyle{myheadings}
\pagestyle{myheadings}

\section{Introduction}

The sandpile model was
introduced by Bak, Tang, and Wiesenfeld (BTW)~\cite{BAK_1}
as a simple example of a slowly driven dissipative system
exhibiting self-organized criticality (SOC). Although today many systems 
with SOC are known, it is considered as the prototype of such 
models, and there is a huge literature devoted to it. 
Its theoretical understanding is crucially related to 
Dhar's discovery of its Abelian structure~\cite{DHAR_2} which
allows exact calculation of many of its properties~\cite{PRIEZ_1,IVASH_1}.
However, a complete analytical determination of the scaling behavior
of avalanches is still lacking.
Several approximation schemes, including
a random walk approach~\cite{OBUKHOV_1},
diffusion-like analogy~\cite{ZHANG_1}, renormalization
group~\cite{PIETRO_3,IVASH_2} and a graph
theory method~\cite{PRIEZ_2}
were proposed, but led to different results.
In addition, computer simulations -- which first had suggested 
simple scaling behavior together with standard finite-size scaling (FSS) --
provide increasing evidence that the avalanche statistics is much 
more complicated. While most recent authors agree upon a breakdown of FFS,
the detailed interpretation of their data is highly controversial 
among different groups~\cite{DEMENECH_1,TEBALDI_1,CHESSA_2,LUEB_6}.

The standard FSS ansatz implies an asymptotic form
\be
    P_a(a,L) \sim a^{-\tau_a}\, p(a/L^{\nu_a})      \label{eq:PaL}
\ee
for the distribution of the number $a$ of toppled sites in an avalanche 
(in other words, its "area"), where $L$ is the size of the lattice, 
$p(z)$ is a universal function, and $\tau_a$ and $\nu_a$ are 
critical exponents. This ansatz implies simple scaling of all
moments of $a$, $\la a^n\ra \sim L^{\sigma_n}$ with 
$\sigma_n=\sigma_0+n\nu_a$. Similar ansatzes should, according to this 
view, hold for the number of topplings $s$ (which differs from $a$ 
because sites can topple more than once in an avalanche) and for the 
radius and duration of avalanches. 
But recent investigations~\cite{DEMENECH_1,TEBALDI_1,CHESSA_2}
show that the two-dimensional BTW model may be characterized
by a multifractal behavior where different moments of $a$ are governed 
by exponents $\sigma_n$ which are not linear in $n$ and are indeed 
not related to each other for different $n$. Different reasons for this 
have been proposed in~\cite{DEMENECH_1,TEBALDI_1} and in~\cite{CHESSA_2}.
Notice that multifractality of avalanches can be proven for certain 
variants of the 1-$d$ sandpile model~\cite{ALI_1}.

\begin{figure}[t]
 \centerline{\psfig{file=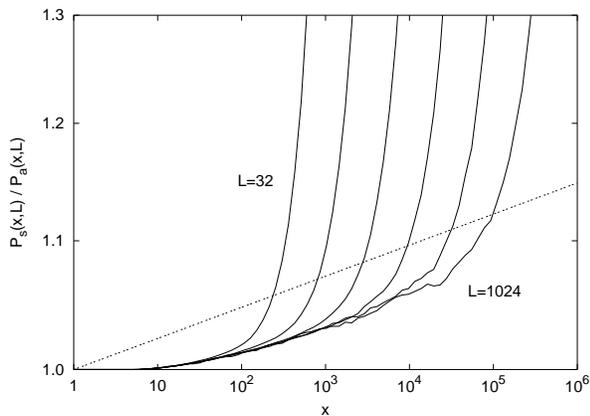,width=8cm,angle=270}}
 \caption{Ratio ${\cal P}_s(x,L)/{\cal P}_a(x,L)$ of the integrated $s$- and 
     $a$-distributions for two-dimensional sandpiles. According 
     to the generally accepted FSS ansatz this should be a power law with 
     exponent $\tau_a-\tau_s$ in the range $0.024$ to $0.08$ in the 
     region where it is independent of $L$. The dashed line is $ x^{0.01}$.
 \label{fig:ratio_Ps_Pa}
 } 
\end{figure}

Deviations from pure power laws had been seen 
already in early simulations, but were usually interpreted as finite 
size effects due to avalanches which touch the boundary of the lattice. 
To illustrate that this is most likely not true, and that there is a 
real problem with simple scaling, we show in Fig.~\ref{fig:ratio_Ps_Pa}
the ratio ${\cal P}_s(x,L)/{\cal P}_a(x,L)$ of the integrated 
distributions ${\cal P}_s(x,L) = \int_x^\infty dx' P_s(x',L)$
for $D=2$ and for different values of $L$. 
In these simulations, cylindrical boundary conditions were used (open 
at $y=\pm L/2$ and periodic at $x=0, L$), and data were collected only 
for avalanches starting at $y=0$. In this way we hope to have minimized 
boundary effects. Also, since we do not make separate fits to $P_s(x,L)$ 
and $P_a(x,L)$, we have none of the uncertainties inherent in such fits.
Due to Eq.~(\ref{eq:PaL}) we would expect this ratio to scale as 
$x^{\tau_a-\tau_s}$ for $x \ll \min\{L^{\nu_a}, L^{\nu_s}\} \approx 
L^2$. According to analytical prediction~\cite{MAJUM_3} 
and recent large scale simulations~\cite{CHESSA_2,LUEB_6,MANNA}, the difference 
$\tau_a-\tau_s$ should be in the range 0.024 to 0.08. 
The behavior seen in 
Fig.~\ref{fig:ratio_Ps_Pa} is rather different. Although the curves 
for different large $L$ perfectly superimpose in a wide range, they 
are in this range not straight at all (as expected for a power law), 
and their average slope in this universal range is much smaller.
Very small differences $\tau_a-\tau_s$ have been seen in several
simulations using small lattices~\cite{CHRIS_2,BIHAM}.
But it still disagrees with our data showing a lack of scaling
even for avalanches which do not reach the boundary of the lattice.
Most other variables show similar deviations from pure 
power laws in $D=2$ when scrutinized closely.

In principle, one can expect that these deviations of scaling can be 
explained by 
assuming that the avalanche boundary advances like a pinned surface in a 
random medium. 
Unfortunately, this interpretation seems untenable. 
As shown in~\cite{IVASH_3} (see also~\cite{GRASS_1}), avalanches 
proceed in distinct {\it waves} of topplings. In each wave, any site 
topples at most once. In the original version of the model waves overlap 
in time, but they can be disentangled by a simple trick~\cite{IVASH_3} 
so that at any time only a single wave propagates.  Therefore, if at all, 
the arguments associated with pinning effects 
should not apply to boundaries of avalanches but to the propagation of 
wave boundaries, and they would suggest that waves show complex 
behavior (notice that boundaries of successive 
waves are not simply related to each other~\cite{PACZUSKI_1}).

But waves {\it do} behave simply, and do show 
simple scaling behavior. This is indeed the main message of the present 
paper. Our results extend analytical results derived 
in~\cite{IVASH_3,DHAR_4,IVASH_4}
and large scale simulations made in~\cite{STELLA_1,PACZUSKI_1}.
While the behavior of avalanches is complex and badly understood 
when typical avalanches are composed of many waves (which is the case 
for $D=2$ and, to a much less degree, for $D=3$), the behavior of 
single waves is simple and well understood. 

In particular, we show in the present paper that FSS works for 
waves of topplings. Since boundary avalanches (i.e.~avalanches which 
start from an unstable boundary site) always consist 
of single waves~\cite{IVASH_3}, it applies also to them. 
Using the spanning
tree representation of waves, the equivalence between spanning trees 
and loop-erased random walks, and rigorous estimations for 
the latter, we determine the critical exponents of their 
probability distributions for all dimensions $D\ge2$. 
We also use the wave statistics to confirm recent
numerical~\cite{LUEB_5} and analytical~\cite{OBUKHOV_1,PRIEZ_99}
predictions for the upper critical dimension of the BTW sandpile 
model. The upper and lower bounds for logarithmic
corrections to scaling for four-dimensional waves
are determined analytically and confirmed numerically.

We discuss the possibility for investigation of avalanche
distributions of the BTW model using the results obtained for 
waves. One of the key characteristics in this study
is the average number $\la n \ra_{a}$ of waves 
in avalanches of a given area $a$. Our simulations show that 
in $D=2$ this number grows, most probably, 
not slower than a power law with the exponent 0.17 (the first 
conjecture of~\cite{MAJUM_3} was 1/6, and an analytical prediction 
of~\cite{PRIEZ_2} is 1/4). This means that multiple
topplings substantially change the scaling behavior of avalanches
in comparison to waves, which was indicated in many previous studies
of the two-dimensional BTW model.

In $D=3$ the fraction of avalanches containing more than one wave
is much less than in two dimensions.
All accurate numerical estimations of the exponent~$\tau_a$
lead to $\tau_a=1.33$ which coincides
with the exact exponent $4/3$ for the wave distribution,
which could mean that the averaged number of waves in an avalanche in
$D=3$ grows not faster than logarithmically. 
On the other hand, considering the numerical estimation
of this number we cannot exclude its slow
polynomial growth with the avalanche size.
Then, the scaling behavior of avalanches could be corrected for the
multiple topplings in large events, similar to the case $D=2$.

Finally, for $D\ge4$ the upper logarithmic bound for the averaged 
number of waves in an avalanche~\cite{PRIEZ_99} implies that the 
distributions of avalanches obey asymptotic behavior with the same 
exponents as for waves.  

This paper is organized as follows. In section~\ref{sec:btw_model} 
we remind basic definitions of the BTW model. 
Section~\ref{sec:waves_trees} is devoted to detailed explanation of 
the construction of waves and their spanning tree representation.
In section~\ref{sec:green_function} we derive
the critical exponents of wave distributions.
In section~\ref{sec:lerw} we discuss analytical results for
the dynamical exponent and fractal dimension of waves.
Results of computer simulations are presented in
section~\ref{sec:numerics}.

\section{The BTW model}
\label{sec:btw_model}

We consider the $D$-dimensional BTW model on a hypercubic lattice
of linear size $L$ in which integer variables $z_i\ge 0$ represent
local energies. One perturbs the system by adding particles at 
randomly chosen sites according to 
\be
    z_i \, \mapsto \, z_i+1\, .             \label{eq:perturbation}
\ee
A site is called unstable if the corresponding energy $z_i$
exceeds the critical value $2D$. An unstable site relaxes, its 
energy is decreased by $2D$ and the energy of the $2D$ nearest 
neighbors (nn) is increased by one:
\be
    z_{i}\;\to\;z_i\,-\,2D                  \label{eq:relaxation_1}
\ee
\be
    z_{\rm nn}\;\to\;z_{\rm nn}\;+\;1.      \label{eq:relaxation_2}
\ee
In this way, the neighboring sites may be activated and
an avalanche of relaxations may proceed.
If a boundary site topples, one or more particles
leave the system. The avalanche of relaxations stops when
all sites are stable again. 

One can introduce in a natural way different kinds of sub-avalanches,
e.g. clusters of sites toppled not less than a given number
of times~\cite{GRASS_1} or waves of topplings~\cite{IVASH_3}.
A relaxation event (an avalanche or sub-avalanche) is characterized
by its size $s$ (total number of topplings),
area $a$ (number of distinct toppled sites),
duration $t$ (number of parallel update steps until stable
configuration is reached),
and its radius $r$ (e.g. the radius of gyration or the 
maximal distance between the origin and a toppled site).
The basic hypothesis of Bak {\it et al.}~\cite{BAK_1} claimed that
in the self-organized critical state the probability
distributions of values $s,a,t,r$ exhibit power-law behavior for 
intermediate values of $x$,
\be
    P_x(x) \sim x^{-\tau_x},                  \label{eq:prob_dist}
\ee
with $x \in \{s,a,t,r\}$.

As we have seen, this hypothesis might not be true for complete 
avalanches, but as we shall see it does hold for waves.
Scaling relations for the exponents $\tau_s, \tau_a, \tau_t$, and
$\tau_r$ can be obtained if one assumes that size, area, 
duration and radius of ``typical" events scale as
powers of each other, for instance
\be
    t \, \sim \, r^{\gamma_{tr}}.                \label{eq:gam_tr}
\ee
Then the transformation law of probability distributions
$P_t(t) \mbox{d}t=P_r(r) \mbox{d}r$ leads to the scaling relation
\be
    {\gamma_{tr}}\;=\;\frac{\tau_r-1}{\tau_t-1}.     \label{eq:gam_tau_tr}
\ee

Again we should warn that there is a crucial assumption underlying 
these relations, namely that conditional distributions $P_x(x|y)$ 
are narrow, and therefore Eq.~(\ref{eq:gam_tr}) holds with small 
deviations for most events. It was proposed in~\cite{CHESSA_2} that 
this might not be justified in $D=2$, and this is indeed the 
main source of problems of this approach. Let us ignore this for the 
moment and proceed nevertheless.

The scaling exponents~$\gamma_{xx'}$ are important for the
description of the extent of avalanches and their
propagation. For instance the exponent $\gamma_{sa}$ indicates
if multiple toppling events are relevant ($\gamma_{sa}>1$)
or irrelevant ($\gamma_{sa}=1$).
The exponent $\gamma_{ar}$ relating the avalanche area to 
its radius $r$~equals the fractal dimension $D_f$ of the avalanche.
Finally, the exponent $\gamma_{tr}$ is usually identified
with the dynamical exponent~$z$.

If Eqs.~(\ref{eq:prob_dist}-\ref{eq:gam_tau_tr}) are applied to 
waves, one has of course $\gamma_{sa}=1$, but $\gamma_{ar}$ 
and $\gamma_{tr}$ are non-trivial.
Our main result states that 
Eqs.~(\ref{eq:prob_dist}-\ref{eq:gam_tau_tr}) 
do indeed apply to waves, together with 
the FSS ansatz Eq.~(\ref{eq:PaL}).

\section{waves of topplings and their spanning tree representation}
\label{sec:waves_trees}

Dhar proved~\cite{DHAR_2} that all stable configurations 
can be classified as either transient or recurrent. The 
former can occur only during an initial transient period, but 
are irrelevant for the infinite time dynamics.
He also formulated the so-called "burning algorithm" which,
on the one hand, allows one to distinguish the recurrent states
among all stable configurations, and, on the other hand, can
be used for constructing a spanning tree representation of any 
given recurrent state. According to this algorithm, which also 
proceeds in discrete time steps, any site $i$ is ``burnt" 
at time $t$ if its energy $z_i$ is larger 
than the number of its unburnt nearest neighbors at time $t-1$.
In a stable configuration, only some of boundary sites can satisfy 
this condition at the first step, they can be interpreted
as origins of ``fire". Then the ``fire" propagates if new sites 
become burnable at the second step.  In this way, we burn the
sites step by step, until no more sites can be burnt. If all
sites of the lattice are burnt, the initial configuration of 
energies is recurrent, otherwise it it transient.

In order to obtain the spanning tree representation of recurrent 
configurations~\cite{MAJUM_3}, 
each burnt site $i$ is connected by a bond to one of the sites 
which had ``set it afire", i.e. which had caused its burning by 
burning itself. If there is more than one such site, one uses an 
arbitrary but definite set of rules where to place this bond. 
In addition, one introduces a new site $\square$ (``sink") and 
connects it to all boundary sites. On these connections, bonds 
are placed to those sites which burn at $t=1$. Then
we can imagine the entire process to start by burning the 
site $\square$ at time $t=0$, and generating a rooted tree 
with root at $\square$. If the state is recurrent, this tree 
spans the entire lattice.

Majumdar and Dhar~\cite{MAJUM_3} also noticed that
the condition for ``toppling" of a site is essentially 
the same as the condition for ``burning": At each step,
the site $i$ topples if its energy $z_i$ is larger than
the number of those of its nearest neighbors which had not 
toppled in the step before. Therefore, 
the burning of a recurrent state is equivalent to a toppling
process initiated ``from the boundary". It implies that 
if we add one particle to every boundary site (two particles 
on each corner, etc.), each site will topple exactly once 
during the ensuing avalanche.

The burning algorithm gives a one-to-one correspondence 
between recurrent states and spanning trees.
This allows one to calculate the total number
of recurrent configurations,
the energy probabilities and the energy-energy
correlation functions~\cite{DHAR_2,PRIEZ_1,IVASH_1,MAJUM_1}.

The spanning tree representation can be constructed also
for a certain class of unstable configurations appearing
during an avalanche.
It was shown in~\cite{IVASH_3} that avalanches in the
BTW model can be decomposed into so called `waves
of topplings'.  According to this construction,
an avalanche is considered as a superposition
of successive sub-avalanches. After 
perturbing the system at a given lattice site~$i$,
one allows it to relax, but prevents 
the site~$i$ temporarily from toppling a second time.
After this first `wave' all sites are again stable except,
possibly, the site $i$. If $i$ is unstable, a second 
wave is initiated by toppling it again. But a possible 
third toppling is again delayed until this wave is finished, 
and when it finally occurs, it triggers the third wave. 
The procedure is repeated until the site~$i$ is stable. Note 
that if the site $i$ is on the boundary, the avalanche stops
after first relaxation and consists of only one wave. More 
generally, if the distance of $i$ to the boundary is $d$, 
then any avalanche starting at $i$ can have at most $d+1$ 
waves.

To obtain the tree representation of a configuration 
following a wave which had started at site $i$, we introduce 
an auxiliary BTW model on a new lattice. In this lattice 
we connect $i$ to the sink $\square$.
Since the site $i$ in the new model is on the boundary,
each avalanche starting at this site will consist of a single
wave. Each avalanche in the auxiliary model corresponds
indeed to a wave of some avalanche in the original model.
Applying the burning algorithm, we can construct a spanning
tree on the auxiliary lattice representing the recurrent
state of the new model. During this burning process, some 
branches of the fire will be independent of site $i$, but 
one branch will first pass from $\square$ to $i$ and then 
propagate further. It is this latter branch which coincides 
with the last wave of topplings in the original BTW model.

\begin{figure}[t]
 \epsfxsize=8.6cm
\epsffile{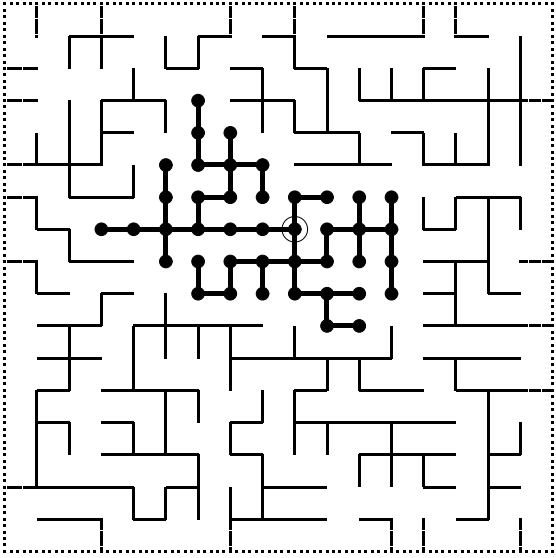}
 \caption{Spanning 2-tree representation of a wave.
          Toppled sites are marked by heavy dots.
          The origin of the wave is marked by a circle.
          The dotted lines indicate the boundary of the system.
          The latter is considered as a single additional site 
          $\square$, so that all non-toppled sites form a 
          single connected tree.
}
 \label{fig:tree_bulk}
\end{figure}

Removing the bond between $i$ and $\square$ we obtain two
trees, one having the root at $i$ and the second at the sink
$\square$. The first tree represents the wave
and the second one corresponds to the sites not toppled in this
wave. The tree with the root $\square$ contains information
about the configuration of stable sites not affected by the wave.
We will call this union of two trees which cover the entire
lattice, a spanning two-component tree (or, simply, a spanning 2-tree).  
According to this tree representation, waves with different
configurations of untoppled sites are counted as different.
An example of a spanning 2-tree is shown in Fig.~\ref{fig:tree_bulk}.

The rigorous proof of the above construction is given in~\cite{IVASH_3}.
A similar decomposition of avalanches into "inverse avalanches"
was proposed by Dhar and Manna~\cite{DHAR_4}.

An important fact concerning the wave statistics should be noted.
Since all recurrent states of the auxiliary BTW model have equal
probability of occurrence~\cite{DHAR_2}, all waves in the original 
BTW model are also equally likely.

\section{Critical exponents and Green functions}
\label{sec:green_function}

Using the graph representation of waves, we can express
their probability distribution by the lattice Green function.
Consider avalanches initiated by adding a particle
at the site $i$ and spanning 2-trees with roots at $i$ and
$\square$ representing waves of these avalanches.
It was proven in~\cite{IVASH_3} that the Green function
$G_{ij}$  is related to the number
${N}_{ (\square)(ij)}$ of spanning 2-trees 
where the site $j$ is in the same component as $i$:
\be
   G_{ij}=\frac{N_{(\square)(ij)}} {N_{(\square)}},
            \label{eq:green_two_rooted}
\ee
where ${N}_{ (\square)}$ is the total number of spanning trees
with the root at the site $\square$.

Consider avalanches initiated by adding a particle at the site 
$i$. Since every recurrent state together with the perturbed 
site $i$ completely defines the relaxation process,
the number of different possible avalanches ${N}^{(a)}_{(i)}$
started at fixed point $i$ equals to the number of recurrent
states (or the number of spanning trees ${N}_{ (\square)}$).
The number ${N}^{\scr(w)}_{(ij)}$ of different waves started at 
the site $i$ and covering the site $j$ corresponds to the 
number ${N}_{(\square)(ij)}$ of 
spanning 2-trees having sites $i,j$ on one of its components.
We can therefore rewrite Eq.~(\ref{eq:green_two_rooted}) as 
\be
    G_{ij}=\frac{N^{\scr(w)}_{(ij)}}{N^{(a)}_{(i)}}.
                                  \label{eq:green_waves}
\ee
Eq.~(\ref{eq:green_waves}) is another formulation of the
known result of Dhar~\cite{DHAR_2} that the expected number of 
topplings at site $j$ due to adding a particle at the site $i$ 
is given by the Green function $G_{ij}$.

Due to uniformness of the wave statistics mentioned at 
the end of section III, the probability  that a wave $W(i)$ 
starting at the site $i$ covers the site $j$ is equal to the
fraction of waves having this property:
\be
     P(j \in W(i))=\frac{N^{\scr(w)}_{(ij)}}{N^{\scr(w)}_{(i)}},
                              \label{eq:prob_waves_01}
\ee
where ${N}^{\scr(w)}_{(i)}$ denotes the total number of
waves starting at $i$.

Combining Eqs.~(\ref{eq:green_waves}) and (\ref{eq:prob_waves_01}), 
we write
\be
    P(j \in W(i))=\frac{N^{\scr(w)}_{(ij)}}
      {N^{(a)}_{(i)}} \frac{N^{(a)}_{(i)}}{N^{\scr(w)}_{(i)}}=
      \frac{G_{ij}}{G_{ii}}=\frac{G(r)}{G(0)}.
                                        \label{eq:prob_waves_02}
\ee
where we use the notation $G(r)$ for the Green function
$G_{ij}$ if the points $i$ and $j$ are separated
by the distance $r$.

On the other hand, this probability can be represented as
\be
    P(j \in W(i)) \; = \; \int_r^\infty P^{\scr(w)}(R)\; 
                  \rho_R(r)\;dR\;,      \label{eq:prob_topp_02}
\ee
where $P^{\scr(w)}(R) $ is the
probability that the linear extent of a wave is $R$, 
and $\rho_R(r)$ denotes the density of sites covered by 
such a wave. The density $\rho_R(r)$ tends to 1 for large $R$ 
if waves are compact and isotropic, and is a function of $r$ 
if waves are fractal. Asymptotically for $R\gg r$, it varies as
\be
    \rho_R(r)\approx \rho(r) \sim r^{d_f-D}.            \label{15}
\ee
where $d_f$ is the fractal dimension of waves and $D$ is
the Euclidean dimension of the lattice.

Suppose that the probability distribution of the wave radius $r$
has a power law asymptotics similar to that for
avalanches [Eq.\,(\ref{eq:prob_dist})],
\be
     P^{\scr(w)}_r(r) \; \sim  \;
     r^{-\tau^{\scr(w)}_r}.
                                   \label{eq:prob_wave_r}
\ee
Then, the probability distribution $
P^{\scr(w)}_{r}(r^\prime > r)$
scales with the exponent $\tau^{\scr(w)}_r-1$.
Using Eq.\,(\ref{15})  we get the asymptotic
behavior of the probability in lhs of Eq.~(\ref{eq:prob_topp_02})
\be
P(j \in W(i)) \; \sim \; r^{-\tau^{\scr(w)}_r+1+d_f-D}.
\label{eq:green_radius}
\ee

The asymptotics  of the bulk Green
function~(see for instance~\cite{JACKSON_ED}) are given by
\be
   G(r) \; \sim \; \left \{
             \begin{array}{ll}
                        \ln{r}   & \quad {\rm for} \; D=2 \\
                                 &             \\
                        r^{2-D}  & \quad {\rm for} \; D>2 \\
             \end{array}
                   \right.
                                  \label{eq:green_asymp_bulk}
\ee
which reveals that the radius exponent 
$\tau^{\scr(w)}_{r}$ for waves is
\be
   \tau^{\scr(w)}_r = d_f-1.
                             \label{eq:tau_radi_wave_bulk}
\ee
Using Eq.\,(\ref{eq:gam_tau_tr}) we can derive 
the exponents of the wave area 
\be
   \tau^{\scr(w)}_a = 2 - \frac{2}{d_f}
                        \label{eq:tau_area_wave_bulk}
\ee
and duration
\be
   \tau^{\scr(w)}_t = 1 + \frac{d_f-2}{z},
                        \label{eq:tau_dura_wave_bulk}
\ee
respectively.

For avalanches started at a distance $b$ from the boundary,
we need the boundary Green functions which can be calculated by
the method of images:
\be
   G({\bf r}) \; \sim \; \left \{
   \begin{array}{ll}
      \ln{|{\bf r}+{\bf b}|} -
      \ln{|{\bf r}-{\bf b}|}  & \quad {\rm for} \; D=2 \\
                              & \\
      |{\bf r}-{\bf b}|^{2-D} -
      |{\bf r}+{\bf b}|^{2-D}  & \quad {\rm for} \; D>2 \\
   \end{array}
   \right.
                     \label{eq:green_asymp_bound_01}
\ee
where ${\bf b}$ is the vector perpendicular to the boundary.
On any ``equipotential" surface $G({\bf r})={\rm const}$ 
characterized by a length scale $\xi$ and a volume 
$a\sim \xi^D$, this boundary Green function scales as 
\be
   G  \sim  b \, \xi^{1-D} \;.       \label{eq:gf_bound}
\ee

If we now replace Eq.~(\ref{15}) (which is appropriate only 
for isotropic cases) by its generalization $\rho \sim 
\xi^{d_f-D}$, we arrive at the exponents for waves 
starting near the boundary:
\be
   \tau^{\scr({\rm boundary})}_r = d_f,
               \label{eq:tau_radi_wave_bound}
\ee
\be
   \tau^{\scr({\rm boundary})}_a = 2 - \frac{1}{d_f},
               \label{eq:tau_area_wave_bound}
\ee
\be
   \tau^{\scr({\rm boundary})}_t = 1 + \frac{d_f-1}{z}
               \label{eq:tau_dura_wave_bound}
\ee
(here and in the following we use superscript
$(w)$ for bulk {\it w}aves and ${(\rm boundary)}$ for boundary waves,
and use symbols without superscripts for avalanches).
We see that both the bulk and boundary wave exponents are
determined by the scaling exponents $d_f$ and $z$. The 
dynamical exponent $z$ can be related to the fractal dimension 
of the ``chemical path" on a spanning tree~\cite{MAJUM_3}
which, in turn, is equivalent to the
dimension of the loop-erased random walk (LERW)~\cite{MAJUM_2}.
As to the fractal dimension of waves $d_f$, it was proven
for all dimensions that a set of untoppled
sites, which are completely surrounded by toppled sites,
corresponds to a forbidden subconfiguration~\cite{CHRIS_2,GRASS_1}.
However, this fact does not prevent waves from being fractal.
They still could display either non-fractal or
fractal behavior depending on the dimensionality $D$.
In the next section, we will show that $d_f$
is also closely related to properties of LERW, more precisely 
to the intersection probability between a LERW and a simple
random walk.

\section{Loop erased random walks, dynamical exponent and
fractal dimension of waves}
\label{sec:lerw}

In this section, we derive analytical estimates for critical 
exponents of waves using their spanning tree representation and
equivalence between a chemical path on a spanning tree and
LERW.

Consider an unrestricted  random walk  on a hypercubic lattice.
The LERW introduced by Lawler~\cite{LAWLER_1} is obtained from 
the simple random walk by deleting all loops along the path.
The chemical path between two sites of a tree
is the unique path along the tree edges connecting these
sites. Majumdar~\cite{MAJUM_2} has shown that the chemical path
on a spanning tree is statistically equivalent  to the LERW,
i.e., the average distance $r$ between the starting point and the
position after $l$ steps scales as $r \sim l^{\nu}$
with the same exponent $\nu$ for both of them.

In $D=2$, the exponent $\nu=4/5$ is known exactly from conformal
field theory~\cite{CONIGLIO_1}. In $D=3$, numerical estimates yield
$\nu\approx0.616$~\cite{GUTTMANN_1,BRADLEY_1}.
In $D = 4$ which is the upper critical dimension
for the LERW, $\nu =1/2$ and the simple scaling 
law has logarithmic corrections~\cite{LAWLER_BOOK}.
For $D > 4$, the scaling behavior of the LERW
and chemical paths is given by the trivial value~$\nu=1/2$,
as the effects of self-intersections become
negligible above the upper critical dimension.

Returning to the BTW model, we notice that sites which topple at a 
given step of wave propagation coincide with sites deleted at the 
same step of the burning process, if it is started at the origin of 
the wave. Since the burning process generates a tree,
there exist a unique path from the root~$i$ of the
tree (the site where the wave is initiated)
to one of the last toppled sites $i_f$  of the wave.
The number of update steps is given by the number of edges in 
this path. Thus, the duration~$t$ of the
wave equals to $l$, the length of the chemical path from~$i$
to~$i_f$ on the tree and, therefore,
the dynamical exponent $\gamma_{tr}$ of waves is given by
$z=1/\nu$.

\begin{figure}[t]
 \epsfxsize=8.6cm
 \epsffile{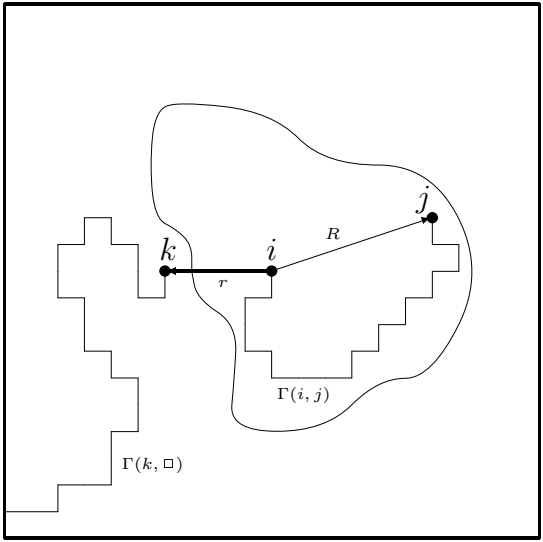}
 \caption{A sketch of a wave initiated at the site~$i$
          and containing the site~$j$.
          The chemical path between the two sites on the
          tree representing the wave is~$\Gamma(i,j)$.
          The second path $\Gamma(k,\Box)$ corresponds to
          a random walk which starts at the site ~$k$
          and escapes the
          path~$\Gamma(i,j)$ until it is trapped at the sink.
         }
 \label{fig:escape_prob}
\end{figure}

In order to find the fractal dimension of waves $d_f$, we use the
proposition proved in~\cite{PRIEZ_99}.
For this we take a site $k$ at distance $r=|k-i|$ from $i$ and 
a site $j$ at distance $R=|j-i|>r$, together with some paths 
$\Gamma(i,j)$ connecting $i$ with $j$ and $\Gamma(k,\square)$
connecting $k$ with the sink $\square$ (see 
Fig.~\ref{fig:escape_prob}). Then the density of sites at 
distance $r$ from $i$, in waves of radius $R>r$ starting at 
site $i$, is given by~\cite{PRIEZ_99}:
\be
   \rho_R(r) = \overline{P_{\rm int}(k \square|i j)},
                           \label{25}
\ee
where $\overline{P_{\rm int}(k \square|ij)}$ is the probability
that $\Gamma(k,\square)$ intersects the path $\Gamma(i,j)$,
averaged over all $j$, all paths from $i$ to $j$, all $k$, 
and all $\Gamma(k,\square)$.

Using the known estimations of the intersection 
probabilities~\cite{LAWLER_BOOK} we can obtain from Eq.~(\ref{25})
the following upper bounds.
For $D>4$, we have
\be
   \rho(r) = \lim_{R\to\infty} \rho_R(r) \leq C_1 r^{4-D},
                \label{26}
\ee
From Eq.~(\ref{15}), we can see that the fractal dimension
of waves $d_f \leq 4$ for all $D \geq 4$.

For $D=4$, the upper bound reads
\be
   \rho_R(r) \leq C_2 \frac{\ln (1+\alpha)}{\ln r}\;,
               \quad \alpha={R^2\over r^2}\;,       \label{27}
\ee
while a lower bound was obtained in~\cite{PRIEZ_99}:
\be
   \rho_R(r) \geq 1 - C_3 \frac{(\ln r)^{1/2}}{(\ln R)^{1/3}}.
                                              \label{28}
\ee
We can see that $\rho_R(r)$ approaches 1 when 
$R \rightarrow \infty,\; r$ fixed. The only fractal dimension
which is consistent with both upper and lower bounds
Eqs.~(\ref{27},\ref{28})
is 4, but there are logarithmic corrections.

For $D<4$ the lower bound Eq.~(\ref{28}) becomes stronger because of
increasing intersection probability $\overline{P_{\rm int}(k \square|i j)}$.
Thus, we conclude that the fractal dimension of waves is
\be
   d_f \; = \; \left \{ \begin{array}{ll}
                 D  & \quad {\rm for} \; D\leq4 \\
                    &         \\
                 4  & \quad {\rm for} \; D>4 \\
               \end{array} \right.
                                          \label{32}
\ee
which means that the upper critical dimension for waves is 4.

Therefore we can calculate from 
Eqs.~(\ref{eq:tau_radi_wave_bulk}-\ref{eq:tau_dura_wave_bulk},
\ref{eq:tau_radi_wave_bound}-\ref{eq:tau_dura_wave_bound})
the exact values of all exponents for bulk and boundary waves
for all dimensions $D\ge 2$, with a single exception. This 
exception is the exponent of the duration distribution in $D=3$, 
for which we need the value of $\nu_{\scr \rm LERW}(D=3)$ which is 
not known exactly.

\section{Comparison with Numerical Simulations}
\label{sec:numerics}
\subsection{D=2}
\label{subsec:two_dim}

In this subsection we present the results of numerical
simulations of bulk and boundary waves in $D=2$. For these 
the standard FSS works well. For any of the observables 
$x=a,t,$ and $r$ it can be written as
\be
   P_x(x,L) \; = \; L^{-\beta_x} \, g_x ( x L^{-\nu_x} ),
                                         \label{eq:fss_simple}
\ee
with $\beta_x = \tau_x \nu_x$~\cite{KADANOFF_1}.

The functions $g_x(z)$ should be universal (i.e., does not 
depend on the type of lattice). But for large 
values of $z$ they do depend 
on the type of boundary conditions (open on all four sides 
or cylindrical, i.e. open in one direction and periodic in 
the other) and on the aspect ratio (square or rectangle 
with sides $L_1\ne L_2$). We verified that the exponents 
were independent of boundary conditions and aspect ratios. 
We verified also that all results were unchanged if we 
threw in the sand grains with non-uniform density, 
provided this density was everywhere non-zero. The latter 
was very useful since it allowed us to obtain much improved 
statistics from either the boundary or the central region.

\begin{figure}[t]
 \epsfysize=6.7cm
 \epsffile{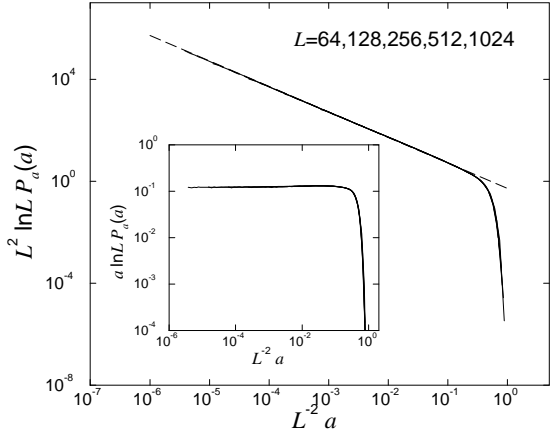}
 \caption{Finite-size scaling plot of the wave area distribution
          $P^{\scr(w)}_a(a)$ for bulk waves in $D=2$.
          The perfect data collapse shows that $d_f=2$ and 
          $\beta_a^{\scr(w)}=2$, as predicted analytically.
          The dashed line represents a power law with exponent
          $\tau^{\scr(w)}_a= 1$. The factor $\ln L$ comes
          from the normalization of the distributions. 
          The inset verifies the scaling 
          ansatz Eq.~(\protect\ref{eq:PaL}).
         }
 \label{fig:a_cfss_wave_2d}
\end{figure}

\begin{figure}[b]
 \epsfysize=6.7cm
 \epsffile{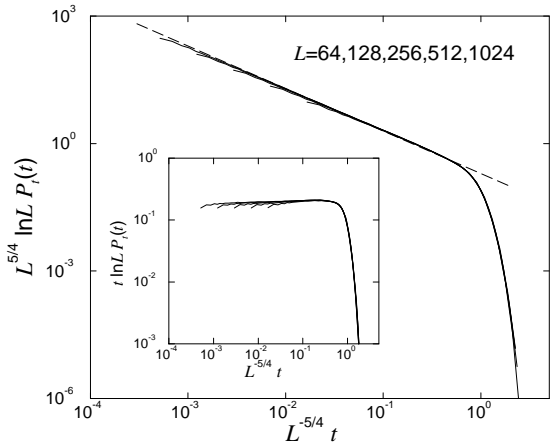}
 \caption{Finite-size scaling plot of the wave duration distribution
          $P^{\scr(w)}_t(t)$ for bulk waves in $D=2$.
          Here, the data collapse is achieved with
          $\beta^{\scr(w)}_t=\nu^{\scr(w)}_t=
          5/4$ which confirms
          again Eq.~(\protect\ref{eq:beta_estimate}).
          The dashed line represents a power law with exponent 
          $\tau^{\scr(w)}_t=1$. Again, the normalization factor
          $\ln L$ is needed for a good data collapse.
         }
 \label{fig:t_cfss_wave_2d}
\end{figure}

Since $d_f=2$ for $D=2$, one has $\nu_a = d_f=2$ and $\nu_r=1$, 
and therefore $\nu_x=\gamma_{xr}$~\cite{LUEB_4}. The results 
of the previous section, together with 
$\nu_t=z=1/\nu_{\scr \rm LERW}=5/4$, give
\be
  \begin{array}{ll}
  \beta^{\scr(w)}_a \; = \; 2\tau^{\scr(w)}_a & = 2, \\
                   & \\
  \beta^{\scr(w)}_t \; = \; z\tau^{\scr(w)}_t & = 5/4, \\
  \end{array}
  \label{eq:beta_estimate}
\end{equation}
for bulk waves and 
\be
  \begin{array}{ll}
  \beta^{\scr({\rm boundary})}_a \; = \; 2\tau^{\scr({\rm boundary})}_a & = 3, \\
                   & \\
  \beta^{\scr({\rm boundary})}_t \; = \; z\tau^{\scr({\rm boundary})}_t & = 9/4, \\
  \end{array}
  \label{eq:beta_estimate1}
\end{equation}
for boundary waves.

\begin{figure}[t]
 \epsfysize=6.7cm
 \epsffile{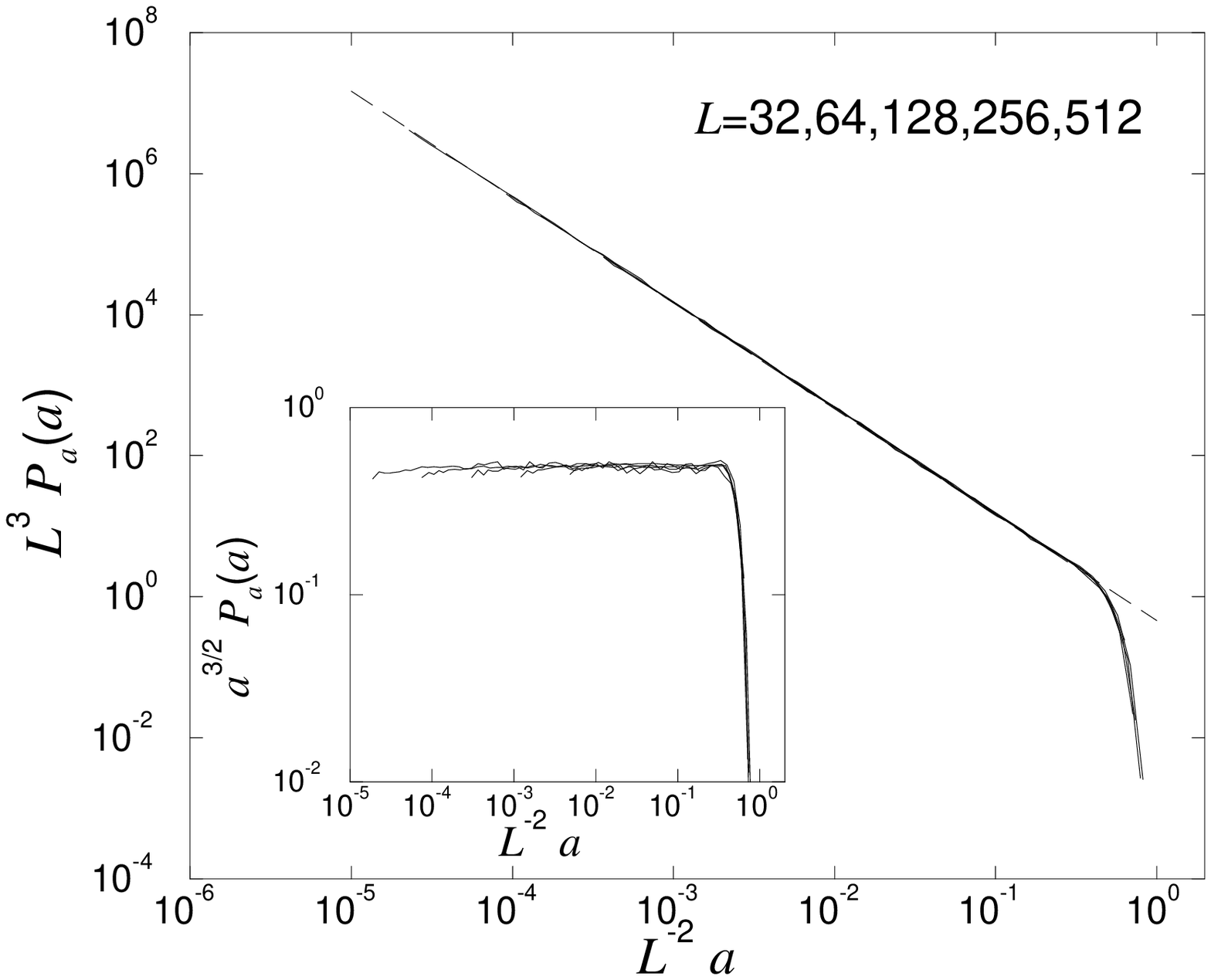}
 \caption{Same as Fig.~\ref{fig:a_cfss_wave_2d}, but for boundary 
          waves in $D=2$. 
          Again the data collapse is obtained with 
          the predicted values $\beta^{\scr({\rm boundary})}_a=3$
          and d$_f=2$. 
          The dashed line corresponds to 
          $\tau^{\scr({\rm boundary})}_a=3/2$,
          as predicted theoretically.
         }
 \label{fig:a_cfss_wave_2d_bound}
\end{figure}

\begin{figure}[b]
 \epsfysize=6.7cm
 \epsffile{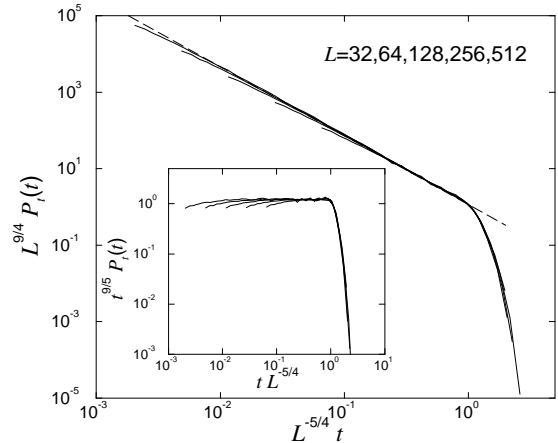}
 \caption{Same as Fig.~\ref{fig:t_cfss_wave_2d}, but for boundary 
          waves in $D=2$. 
          Again the data collapse is obtained with 
          the predicted values $\beta^{\scr({\rm boundary})}_t=9/4$ and 
          $\nu^{\scr(w)}_t=5/4$. 
          The dashed line corresponds to 
          $\tau^{\scr({\rm boundary})}_t=9/5$, 
          as predicted theoretically.
         }
 \label{fig:t_cfss_wave_2d_bound}
\end{figure}

The finite-size scaling plot for the area distribution
$P^{\scr(w)}_a(a)$ of bulk waves in $D=2$
is shown on Fig.~\ref{fig:a_cfss_wave_2d}.
In the inset of this Figure, as well as in the insets
of plots for other distributions of waves, we 
show the collapses according to ansatz of Eq.~(\ref{eq:PaL}).   
Taking $\beta^{\scr(w)}_a=2$ and $d_f=2$ we see 
a perfect data collapse. 
The finite-size scaling ansatz of the duration 
distribution is plotted in Fig.~\ref{fig:t_cfss_wave_2d}. 
These data confirm that waves 
are not fractal, and that their duration is as predicted by 
the correspondence with spanning trees and loop-erased random 
walks.

As was mentioned above, the avalanches started at the boundary 
consist of a single wave, so for this type of avalanches
the distribution of avalanches coincides with that of waves.
The finite-size scaling plots for the
area and duration distributions for boundary waves
are shown in Fig.~\ref{fig:a_cfss_wave_2d_bound} and 
Fig.~\ref{fig:t_cfss_wave_2d_bound}, respectively.
Again all predictions are verified, in particular we see that 
$d_f=2$, i.e. also boundary waves are not fractal in $D=2$.

\begin{figure}[t]
 \epsfysize=6.6cm
 \epsffile{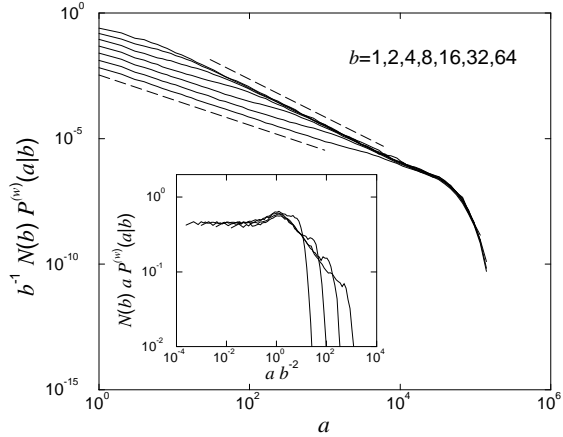}
 \caption{Area distributions of waves initiated 
          at different distances~$b$ from the boundary
          for a system of size $256\times 1024$ with cylindrical
          boundary conditions.
          Each curve is averaged over approximately
          $10^6$ avalanches.
          The dashed lines correspond to the bulk
          ($\tau^{\scr(w)}_a=1$) and boundary 
          ($\tau^{\scr({\rm boundary})}_a=3/2$)
          scaling behavior, respectively. The distributions
          are multiplied by $b^{-1}N(b), N(b)=\int P_a^{(w)}(a|b)da$ 
          in order to have the curves 
          collapsing for large $a$. 
          Inset: the rescaled distributions for
          $b=8,16,32,64$ demonstrate that the crossover from
          bulk to boundary behavior takes place at values of
          area of order $b^2$.
 \label{fig:cross_over_2d}}
\end{figure}

Finally, let us consider avalanches starting at a finite 
distance from the boundary. The crossover from the boundary to
bulk behavior of the wave distribution should be described by 
Eq.~(\ref{eq:green_asymp_bound_01}). More precisely, the 
equipotential surfaces in $D=2$ are circles~\cite{JACKSON_ED} 
with radius $\xi$
and $G \sim \ln [(b^2/\xi^2+1)^{1/2}-b/\xi]$. In the scaling 
region where $\rho(\xi)=1$, we have therefore 
$P^{\scr(w)}_a(a|b) = -(da/d\xi)^{-1} dG/d\xi$ which gives
\be
   P^{\scr(w)}_a(a|b) \sim \; {b\over a\sqrt{a+\pi b^2}} \sim \left \{
             \begin{array}{ll}
                     b/a^{3/2}   & \; {\rm for} \; a > \pi b^2 \\
                                 &             \\
                     1/a         & \; {\rm for} \; a < \pi b^2 \\
             \end{array}
                   \right.
                                  \label{eq:Pa-boundary-2d}
\ee
To check this, we simulated the BTW model with cylindrical boundary 
conditions, and collected data for waves started at distance $b$
from the open boundary. 
The results are plotted in Fig.~\ref{fig:cross_over_2d}. 
For small $a$ we see indeed the bulk 
behavior which crosses over to the boundary behavior $a^{-3/2}$ 
for $b^2 < a < L^2$. In the latter region we also see clearly 
the linear dependence on $b$. 

The above shows that our understanding of waves in the 
two-dimensional BTW model is basically complete. In contrast, and 
in spite of numerous efforts, the scaling behavior of
{\it avalanches} in the two-dimensional BTW model
is still an open problem. This is due to multiple 
topplings. The average number of waves in an avalanche 
scales as~\cite{DHAR_2}
\be
   \la n \ra \sim \ln L \;.   \label{eq:n_w}
\ee
There are also several results known 
about correlations in the sizes of successive 
waves~\cite{PRIEZ_2,PACZUSKI_1,KTITAREV_1,IVASH_unpub}. 
Nevertheless, even most 
basic questions such as the distribution of $n$ or 
the dependence of $n$ on the area $a$ are not yet solved.

\begin{figure}[t]
 \epsfysize=6.1cm
 \epsffile{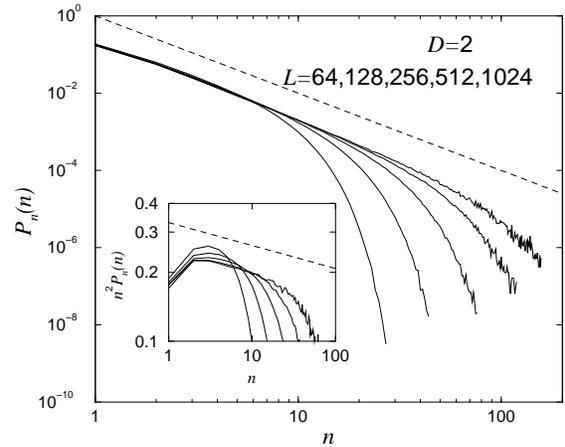}
 \caption{Probability distribution of the number of
          waves $n$ in an avalanche for $D=2$.
          The dashed line corresponds to the 
          power-law behavior $P_n(n)\sim n^{-2}$ as
          predicted in~\protect\cite{MAJUM_3}.
          The data were collected from 
          avalanches initiated at the center region of the
          square lattice with cylindrical boundaries.
          The inset shows the same data multiplied by $n^2$
          There, the dashed line is $\propto n^{-0.1}$.
 \label{fig:n_dist_2d}}
\end{figure}

\begin{figure}[b]
 \epsfysize=6.1cm
 \epsffile{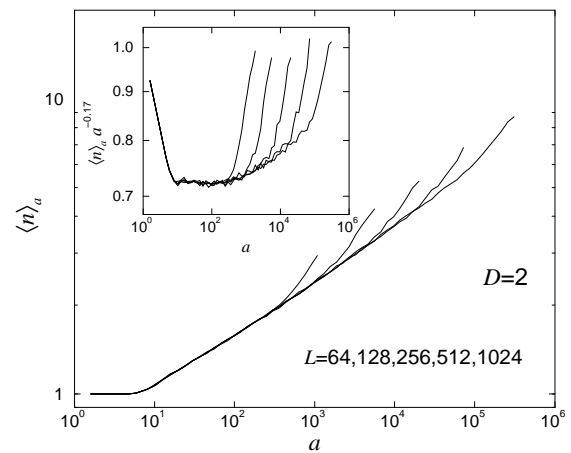}
 \caption{Average number of waves~$\langle n \rangle_a$
          as a function of the avalanche area~$a$ for
          $D=3$ and various system sizes $L$. 
          Although the main figure looks straight at first,
          the inset displaying the rescaled average
          shows significant deviations from the
          assumed pure power-law behavior~\protect\cite{MAJUM_3}. 
          The data were collected from non-dissipative
          avalanches initiated at the center region of the
          square lattice with open boundaries. Thus the curvature
          seen in the inset cannot come from avalanches reaching
          the boundary.
 \label{fig:gam_n_a_2d}}
\end{figure}

Equation (\ref{eq:n_w}) would be most easily explained if 
$P_n(n)$ were simply $\sim 1/n^2$. Present data seem to agree 
with this for the largest lattices (Fig.~\ref{fig:n_dist_2d}), 
but actually the data are better 
fited with a power $1/n^{2.1}$ than with $1/n^2$ (see 
inset).
Similar results are obtained for 
$\la n\ra_a$, the average number of waves in avalanches 
with fixed $a$ (Fig.~\ref{fig:gam_n_a_2d}). 
Although they seem to scale like a power of $a$, 
as assumed in~\cite{MAJUM_3}, a closer study shows significant 
deviations which seem hard to explain as finite size effects.

There are several recent papers~\cite{DEMENECH_1,DROSSEL_1} 
which try to explain these problems by unexpected features 
of avalanches which reach the boundary. But data such as 
those shown in Figs.~\ref{fig:ratio_Ps_Pa},\ref{fig:gam_n_a_2d} indicate that 
there are already problems with avalanches which do not 
reach the boundary.

\subsection{D=3}
\label{subsec:three_dim}

\begin{figure}[b]
 \epsfysize=6.8cm
 \epsffile{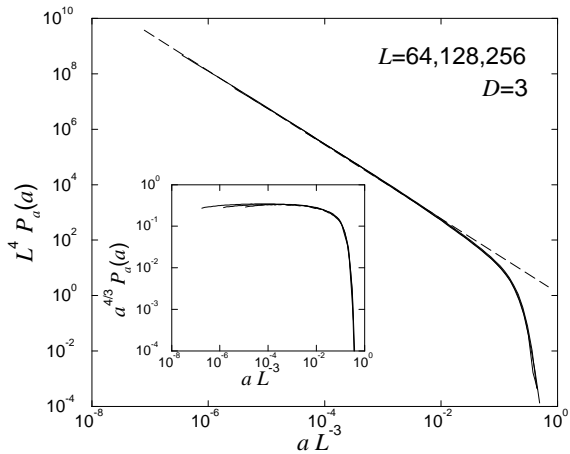}
 \caption{Finite-size scaling plots of the area 
          distribution $P^{\scr(w)}_a(a)$ for bulk 
          avalanches in $D=3$.
          Assuming compact avalanches ($d_f=3$) we get
          good data collapses and the resulting $\tau^{(\scr w)}_a$ 
          exponent agrees with the 
          theoretical prediction (dashed lines).
         }
 \label{fig:a_cfss_wave_3d}
\end{figure}

\begin{figure}[t]
 \epsfysize=6.7cm
 \epsffile{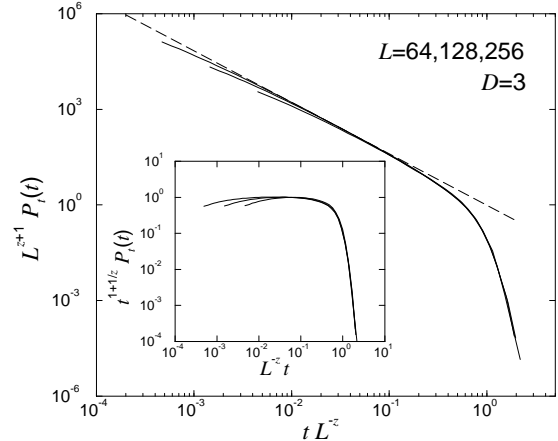}
 \caption{Finite-size scaling plots of the duration 
          distribution $P^{\scr (w)}_t(t)$ for bulk 
          avalanches in $D=3$.
          Using compact avalanches ($z=1/\nu_{\scr \rm LERW}$) we get
          good data collapses and the resulting $\tau^{(\scr w)}_t$ 
          exponent agrees with the 
          theoretical prediction (dashed lines).
         }
 \label{fig:t_cfss_wave_3d}
\end{figure}

\begin{figure}[b]
 \epsfysize=6.8cm
 \epsffile{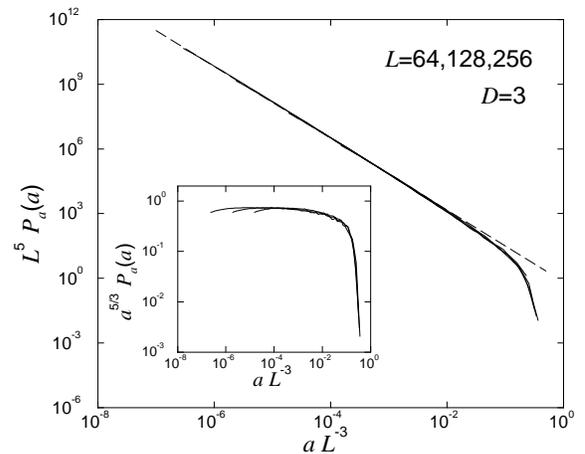}
\caption{Finite-size scaling plots of the area 
          distribution for boundary waves in $D=3$. 
          The data collapses confirm again $d_f=3$, 
          and the dashed line demonstrates the agreement with 
          the predicted value for $\tau_a^{({\rm boundary})}$.
         }
 \label{fig:a_cfss_wave_3d_bound}
\end{figure}

\begin{figure}[t]
 \epsfysize=6.8cm
 \epsffile{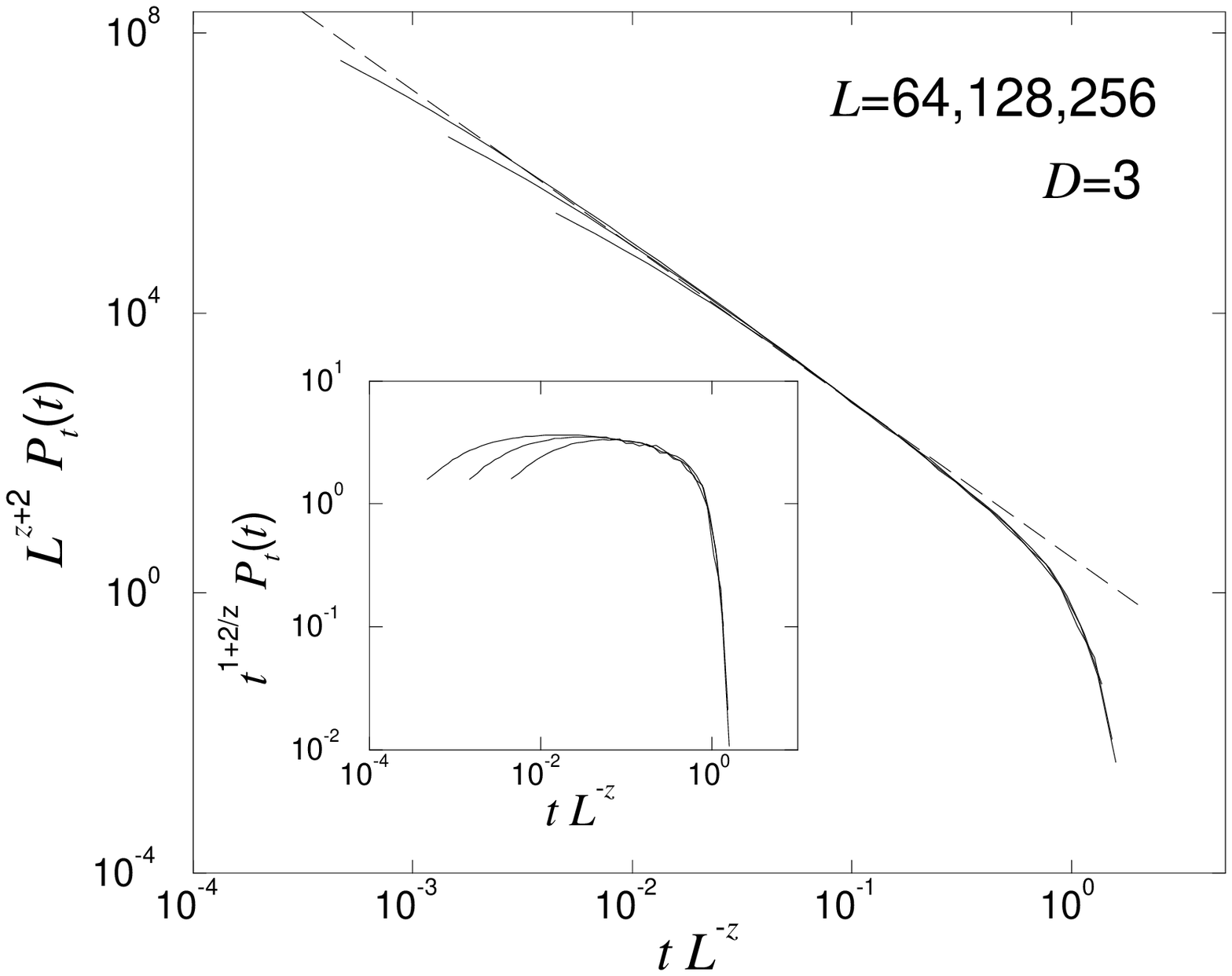}
\caption{Finite-size scaling plot of the duration 
          distributions for boundary waves in $D=3$. 
          The data collapses confirm again $z=1/\nu_{\scr \rm LERW}$, 
          and the dashed line demonstrates the agreement with 
          the predicted value for $\tau_t^{({\rm boundary})}$.
         }
 \label{fig:t_cfss_wave_3d_bound}
\end{figure}

Simulations of waves of the BTW model in $D=3$
also give good agreement with our analytical results
of Sections \ref{sec:green_function} and \ref{sec:lerw}. 
For bulk waves we now have $\beta^{\scr(w)}_a=4,\; 
\tau^{\scr(w)}_a = 4/3,\; \beta^{\scr(w)}_t = 1/\nu_{\scr \rm LERW}+1 \approx 
2.623,$ and $\tau^{\scr(w)}_t = 1+\nu_{\scr \rm LERW} \approx 1.616$. 
For boundary waves, the corresponding numbers are 
$\beta^{\scr({\rm boundary})}_a=5,\; \tau^{\scr({\rm boundary})}_a = 5/3,\;
\beta^{\scr({\rm boundary})}_t = 1/\nu_{\scr \rm LERW}+2\approx 3.623$, and 
$\tau^{\scr({\rm boundary})}_t = 1+2\nu_{\scr \rm LERW} \approx 2.232$.
For these $\beta$ values, the data collapses of bulk
(Fig.~\ref{fig:a_cfss_wave_3d},\ref{fig:t_cfss_wave_3d}) and boundary 
(Fig.~\ref{fig:a_cfss_wave_3d_bound},\ref{fig:t_cfss_wave_3d_bound}) 
waves are perfect. 
They confirm also the analytical predictions for the $\tau$ exponents, 
verifying in particular that the waves have fractal dimension 
$d_f=3$.

\begin{figure}[h]
 \epsfxsize=7.2cm
 \epsffile{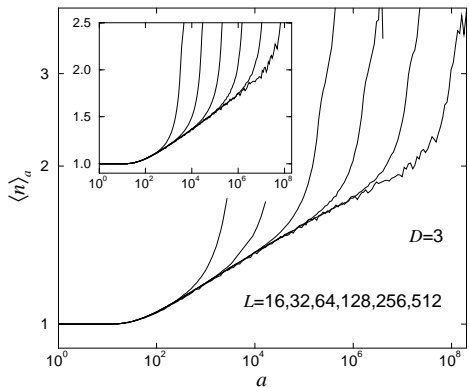}
 \caption{Average number of waves~$\langle n \rangle_a$
          as a function of the avalanche area~$a$ for
          $D=3$ and various values of the system size $L$. In order to 
          minimize finite size effects, cylindrical 
          boundary conditions with one open and two periodic 
          directions were used and one half of all sand grains 
          were thrown onto the central planes $y=L/2$ and 
          $y=L/2+1$. The y-axis 
          is logarithmic in the main plot, and linear in the 
          inset. Neither way of plotting gives perfectly 
          straight lines in the region where that data for 
          different $L$ collapse. Although the main figure 
          looks more straight at first sight, a more careful 
          inspection shows a slight downward curvature.
 \label{fig:gam_n_s_3d}}
\end{figure}

Due to the rarity of multiple topplings, avalanche distributions 
coincide within the displayed accuracy with wave distributions, 
when plotted as in Fig.~\ref{fig:a_cfss_wave_3d} and
Fig.~\ref{fig:t_cfss_wave_3d}. 
In order to show significant results for multiple topplings, we have 
to present the data differently. In Fig.~\ref{fig:gam_n_s_3d} 
we plotted the average number of waves at fixed $a$,
$\la n\ra_a$, against $\log a$. Neither using a logarithmic 
scale for $\la n\ra_a$ (main figure) nor a linear scale 
(inset) give perfectly straight lines. Thus the data can
be interpreted either as a power law with a very small 
exponent,
\be
   \langle n \rangle_a \sim a^\alpha \;\;,\quad \alpha \approx 0.06
                                \label{eq:n_s_D_3}
\ee
or as a logarithmic growth.


In the latter case, we would of course have 
$\tau^{\scr(w)}_a=\tau_a=\tau_s$. In the opposite case of 
a power law with exponent $\alpha \approx 0.06$ we can give 
crude estimates for the differences between these $\tau$ 
exponents, using some heuristic assumptions. 

The first assumption is that different waves 
in the same avalanche involve essentially the same sites. 
If this is true, we should have 
$P^{\scr(w)}_a(a)da \approx \la n \ra_a P_a(a)da $.
Using this together with Eq.~(\ref{eq:n_s_D_3}), we  
find $\tau_a = 4/3 + \alpha \approx 1.39$. Since the basic 
assumption here is most likely not justified, this is 
only a very crude (and most likely too large, in particular since 
the growth of $\la n \ra_a P_a(a)da $ could be logarithmic) 
estimate for 
the difference between $\tau_a$ and $\tau^{\scr(w)}_a$.

An estimate for the difference between $\tau_a$ and $\tau_s$ 
is obtained as follows. An upper bound for the size
of an avalanche of area $a$ containing $n$ waves is $s < na$.
Therefore, the assumption $s \sim \la n \ra_a a$ leads to the
maximal difference between the area and size exponents. Using
Eq.~(\ref{eq:n_s_D_3}) we get $a \gtrsim s^{1/(1+\alpha)}$. Then,
Eq.~(\ref{eq:gam_tau_tr}) gives $(\tau_s-1)\geq (\tau_a-1)/(1+\alpha)$
and we can conclude that
\be
\tau_a-\tau_s \leq \frac{\alpha}{1+\alpha} (\tau_a-1) \leq 0.02.
\label{eq:estimate_tau_s}
\ee

A direct verification of such slight differences between the 
$\tau$ exponents could be tried by plotting ratios of the 
distributions, as was done in Fig.~\ref{fig:ratio_Ps_Pa} for $D=2$. 
We do not show any such ratio here, since they are all 
very close to 1 in the region where the distributions 
should follow power laws, and the deviations from 1 seem 
not to be simple powers.
\vfill

\subsection{D=4}

As the dimension of the BTW model increases, multiple 
toppling events in avalanches become more and more rare. 
For $D\ge 4$ it was shown in~\cite{PRIEZ_99} that 
$\la n \ra_a$ grows not faster than logarithmically, 
i.e. $\alpha=0$. As we already mentioned in the previous 
subsection, this means $\tau_a^{\scr(w)}=\tau_a=\tau_s$,
i.e., the scaling behavior of waves and avalanches in $D=4$ is
characterized by the same exponents and scaling functions.

At the upper critical dimension $D_{\rm u}=4$ logarithmic 
corrections to the simple scaling are essential.
The probability distributions of the radius, duration, area
and the scaling relations between them are expected to have
the form (cf.~\cite{LUEB_5})
\be
   P_r(r) \sim \frac{(\ln r)^{x_r}}{r^3}, \;
   P_t(t) \sim \frac{(\ln t)^{x_t}}{t^2}, \;
   P_a(a) \sim \frac{(\ln a)^{x_a}}{a^{3/2}},
                        \label{eq:prob_dist_4d}
\ee
and
\be
   a \; \sim \; \frac{r^4}{(\ln r)^{N_a}}, \quad
   t \; \sim \; \frac{r^2}{(\ln r)^{N_t}},
                        \label{t}
\ee
respectively.

The  exponents of logarithmic corrections
$x_r$, $x_a$, $x_t$, $N_a$, and $N_t$ obey
the scaling relations~\cite{LUEB_5}
\be
   x_r=x_a + N_a/2, \quad x_r=x_t + N_t
\label{eq:scal_rel_xr_xa_xt}
\ee
which follow straightforwardly from Eqs.~(\ref{eq:prob_dist_4d}) 
and (\ref{t}).

Using arguments similar to those at the end of the previous subsection,
we obtain an inequality for the exponents of logarithmic corrections
for waves and avalanches
\be
   x_a^{\scr(w)} = x_a \le x_s.
\label{x}
\ee

This allows us to compare below analytical estimations for waves
with numerical results for avalanches.

\begin{figure}[t]
 \epsfysize=6.8cm
  \epsffile{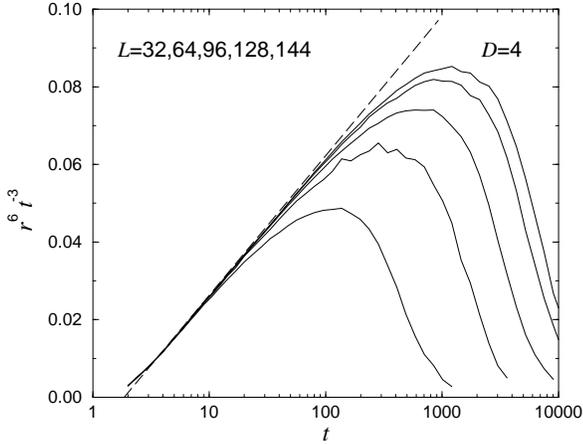}
 \caption{The gyration radius of avalanches as a function of
          their duration.
          We plot the sixth power of the average rescaled gyration radius,
          $[r_g^2 t^{-1}]^3$, in a logarithmic diagram, since this should result 
          in a straight line according to Eq.~(\protect\ref{eq:lerw_scal_log}).
          Such a linear regime is indeed observed (dashed line; its
          slope and intercept are not predicted by theory), 
          and it increases with the system size~$L$.
 \label{fig:gam_r_t_4d}}
\end{figure}

It follows from Lawler's results~\cite{LAWLER_BOOK} discussed in 
Sect.~\ref{sec:lerw} that 
$N_t=1/3$ exactly. But as with all logarithmic corrections, 
a numerical verification is not easy. The main reason is that 
the logarithms are never very much larger than 1, even for the 
largest simulations. Therefore the next-to-leading terms 
(which are typically suppressed by powers of the same 
logarithms) are in general not negligible. In view of this,
the disagreement with recent simulations~\cite{LUEB_5} which 
had suggested $N_t \approx 1/2$ should not be taken serious.
Data for the mean squared radius of avalanches with fixed duration $t$ 
are shown in Fig.~\ref{fig:gam_r_t_4d}. More precisely, since we 
expect 
\be
   {\la r\ra_t^2 \over t} \sim (\ln t)^{1/3},
          \label{eq:lerw_scal_log}
\ee
we plotted $[\la r\ra_t^2 /t]^3$ against $\ln t$. Apart from very
large $t$ when the finiteness of the lattice makes itself seen, 
we observe essentially a straight line (which is a bit fortuitious
since there are also $1/t$ corrections which are 
important for small $t$). At the same time, a power
law dependence $\la r\ra_t^2 \sim t^{2\nu}$ with $\nu>1/2$, as 
would be expected if $D_c>4$, seems ruled out.

\begin{figure}[t]
 \epsfysize=6.8cm
\epsffile{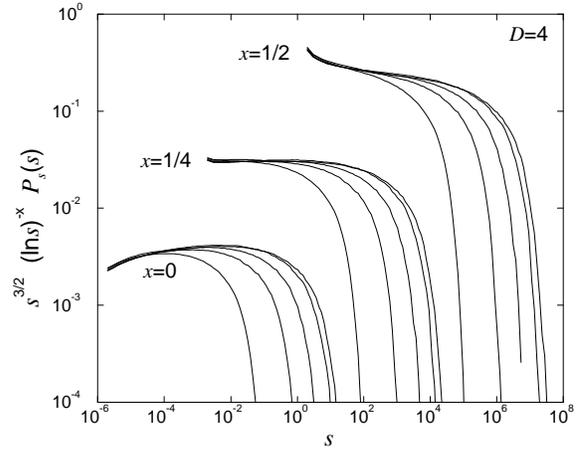}
 \caption{Area distributions for avalanches in $D=4$. 
          In order to render the plots more significant, first of all 
          the dominant $s$-dependence was removed by multiplying 
          with $s^{3/2}$. Then, in order to check whether the remaining 
          $s$-dependence in the scaling region is compatible 
          with logarithmic corrections as proposed in 
          Eq.~(\protect\ref{eq:prob_dist_4d}), we also 
          divided by powers $(\ln s)^x$ and shifted the resulting curves 
          horizontally and vertically in order to avoid overlaps.
          The best agreement is found with $x\approx 0.25$.
         }
 \label{fig:a_fss_4d_log}
\end{figure}

For the other exponents $x_r, x_a, x_t, N_a$ we can only give 
inequalities from the analytical results of section~\ref{sec:lerw}.
The upper bound Eq.~(\ref{27}) for 
the density of waves leads to the relation
\be
   \rho(r) \; \sim \; (\ln r)^{-\delta}, \quad \delta \ge 1.
\label{rho}
\ee
Using this asymptotics and Eq.~(\ref{eq:prob_waves_02}) we get
\be
   P_r^{\scr(w)}(r) \; \sim \; G(r)/\rho(r) \; 
                   \sim \; \frac{(\ln r)^\delta}{r^3},
\label{P(r)_alpha}
\ee
which gives $x_r=\delta$.
The area~$a$ of a wave scales in leading order as
\be
  a \; \sim \; \int_1^r \rho(r')\, r'^3 \, dr' \sim \frac{r^4}{(\ln r)^\delta},
\label{a_alpha}
\ee
giving $N_a=\delta$.
Hence, from Eq.~(\ref{eq:scal_rel_xr_xa_xt}) we have
\be
    x_a \; = \; \delta /2 \; \ge \; 1/2.             \label{x_a}
\ee

In order to verify these predictions -- and to verify, in the 
first place, that deviations from power laws with the mean 
field exponents $\tau_r=3,\tau_t=2, \tau_a=3/2$ cannot be 
eliminated by changing these exponents -- we performed extensive 
simulations.
Numerical data of the size distribution $P_s(s)$ are shown 
in Fig.~\ref{fig:a_fss_4d_log}. 
Lattice sizes range from $L=32$ to $L=144$. 
After multiplying with $s^{3/2}$, we see 
indeed no indication that the remaining $s$-dependence follows 
a clean power law. In order to find the expected logarithmic 
corrections, we multiplied these data by $(\ln s)^{x_a}$, with 
several trial values for the exponent $x_a$. Taken at face 
value, this would give best fits with $x_a \approx 0.25$. In 
view of inequality~(\ref{eq:scal_rel_xr_xa_xt}) and of the 
difficulties in obtaining correct logarithmic corrections 
mentioned above, we propose that indeed $x_a=1/2$. From 
the relations Eq.~(\ref{eq:scal_rel_xr_xa_xt})
we get then $x_r=N_a=1$ and $x_t=2/3$.

\section{Conclusions}

We have studied probability distributions of waves of topplings
in the BTW model on $D$-dimensional rectangular lattices for
$D\ge2$. We have proved analytically that waves as well as
boundary avalanches do exhibit critical
behavior and that their probability distributions display power 
law asymptotics. We have derived exact values of critical 
exponents of these distributions. We have proven analytically 
that the upper critical dimension of the BTW model is $D_{\rm u}=4$, 
showing that previously observed deviations from mean field 
behavior at $D=4$~\cite{CHRIS_2,CHESSA_1} 
are due to logarithmic corrections.
All these results have been confirmed by extensive 
numerical simulations. During these simulations we have also 
verified that wave distributions follow the standard finite size 
scaling ansatz.
The exponent of the leading logarithmic correction to the 
distribution of avalanche life times (or, more precisely, lifetimes 
of waves) has been derived exactly from known asymptotics of 
loop-erased random walks. Estimations are given 
for the exponents of the logarithmic corrections to the other 
distributions. 

We therefore have now a rather complete picture of the dynamics 
of single waves in the BTW model for all dimensions. For $D\ge 4$ 
this means that we also understand avalanche dynamics, since 
multiple topplings are so rare that they can be neglected. For 
$D=2$ the latter is certainly not true, and our understanding 
of avalanche dynamics is still incomplete. 
For $D=3$, finally, multiple topplings represent a small 
but not negligible effect, and we have hope that the 
problem will be solved soon.

\acknowledgements
The authors thank E.\,V.~Ivashkevich for very helpful remarks.
DVK acknowledges the financial support from the Alexander von 
Humboldt Foundation and the kind hospitality of the Computational 
Physics group in Duisburg University,
where part of this work was done.  
VBP was supported by RFBR through Grant No 99-01-00882
and by the Heisenberg-Landau program. 
He also thanks NIC, Forschungszentrum J\"ulich for 
the support and hospitality.


\begin{thebibliography}{10}

\bibitem{BAK_1}
P.~Bak, C.~Tang, and K.~Wiesenfeld, Phys.~Rev.~Lett. {\bf 59},  381  (1987).

\bibitem{DHAR_2}
D.~Dhar, Phys.~Rev.~Lett.~{\bf 64},  1613  (1990).

\bibitem{PRIEZ_1}
{V.\,B.~Priezzhev}, J.~Stat.~Phys. {\bf 74},  955  (1994).

\bibitem{IVASH_1}
{E.\,V.~Ivashkevich}, J.~Phys.~A {\bf 27},  3643  (1994).

\bibitem{OBUKHOV_1}
{S.\,P.~Obukhov},  in  {\it
  Random Fluctuations and Pattern Growth}, edited by H.~E. Stanley and
  N.~Ostrowsky, NATO ASI Series E:~Applied Sciences Vol.~157 (Kluwer,
  Dordrecht, 1988).

\bibitem{ZHANG_1}
Y.-C. Zhang, Phys.~Rev.~Lett.~{\bf 63},  470  (1989).

\bibitem{PIETRO_3}
L.~Pietronero, A.~Vespignami, and S.~Zapperi, Phys. Rev.~Lett.~{\bf 72},  1690
  (1994).

\bibitem{IVASH_2}
{E.\,V.~Ivashkevich}, Phys.~Rev.~Lett.~{\bf 76},  3368  (1996).

\bibitem{PRIEZ_2}
{V.\,B.~Priezzhev}, {D.\,V.~Ktitarev}, and {E.\,V.~Ivashkevich}, 
  Phys.~Rev.~Lett. {\bf 76},  2093  (1994).

\bibitem{DEMENECH_1}
M. {De\,Menech}, {A.\,L.~Stella}, and C. Tebaldi, Phys.~Rev.~E {\bf 58},  2677
  (1998).

\bibitem{TEBALDI_1} 
C. Tebaldi, M. {De\,Menech}, and {A.\,L.~Stella},
preprint cond-mat/9903270 (1999).

\bibitem{CHESSA_2} A.~Chessa, H.\,E.~Stanley, A.~Vespignani, and S.~Zapperi, 
            preprint cond-mat/980263 (1998).

\bibitem{LUEB_6} S.~L\"ubeck and K.\,D.~Usadel, Phys.~Rev.~E {\bf 55}, 4095 (1997)

\bibitem{ALI_1}
{A.\,A.~Ali} and D.~Dhar, Phys.~Rev.~E {\bf 52},  4804  (1995).

\bibitem{MAJUM_3}
{S.\,N.~Majumdar} and D.~Dhar, Physica A {\bf 185},  129  (1992).

\bibitem{MANNA}
{S.\,S. Manna}, Physica (Amsterdam) {\bf 179A}, 249 (1991).

\bibitem{CHRIS_2}
K.~Christensen and Z.~Olami, Phys.~Rev.~E {\bf 48},  3361  (1993).

\bibitem{BIHAM}
E.~Milshtein, O.~Biham, and S.~Solomon, Phys.~Rev.~E {\bf 58},  303  (1998).

\bibitem{IVASH_3}
{E.\,V.~Ivashkevich}, {D.\,V.~Ktitarev}, and {V.\,B.~Priezzhev}, Physica A {\bf
  209},  347  (1994).

\bibitem{GRASS_1}
P.~Grassberger and S.\,S.~Manna, J.~Phys.~(France) {\bf 51},  1077  (1990).

\bibitem{PACZUSKI_1}
M.~Paczuski and S.~Boettcher, Phys.~Rev.~E {\bf 56},  R3745  (1997).

\bibitem{DHAR_4}
D.~Dhar and {S.\,S.~Manna}, Phys.~Rev.~E {\bf 49},  2684  (1994).

\bibitem{IVASH_4}
{E.\,V.~Ivashkevich}, {D.\,V.~Ktitarev}, and {V.\,B.~Priezzhev}, J.~Phys.~A
  {\bf 27},  L585  (1994).

\bibitem{STELLA_1}
{A.\,L.~Stella}, C.~Tebaldi, and G.~Caldarelli, Phys.~Rev.~E {\bf 52},  72
  (1995).

\bibitem{LUEB_5}
S. L{\protect\"u}beck, Phys.~Rev.~E {\bf 58},  2957  (1998).

\bibitem{PRIEZ_99}
{V.\,B.~Priezzhev}, preprint cond-mat/9904054 (1999).

\bibitem{MAJUM_1}
{S.\,N.~Majumdar} and D.~Dhar, J.~Phys.~A {\bf 24},  L357  (1991).

\bibitem{JACKSON_ED}
{J.\,D.~Jackson}, {\em Classical electrodynamics} (Wiley, New York, 1975).

\bibitem{MAJUM_2}
{S.\,N.~Majumdar}, Phys.~Rev.~Lett.~{\bf 68},  2329  (1992).

\bibitem{LAWLER_1}
{G.\,F.~Lawler}, Duke Math.~J.~{\bf 47},  655  (1980).

\bibitem{CONIGLIO_1}
A.~Coniglio, Phys.~Rev.~Lett.~{\bf 62},  3054  (1989).

\bibitem{GUTTMANN_1}
{A.\,J.~Guttmann} and {R.\,J.~Bursill}, J.~Stat.~Phys.~{\bf 59},  1  (1990).

\bibitem{BRADLEY_1}
{R.\,E.~Bradley} and S.~Windwer, Phys.~Rev.~E {\bf 51},  241  (1995).

\bibitem{LAWLER_BOOK}
{G.\,F.~Lawler}, {\em Intersection of Random Walks} ({Birkh\"auser}, Boston,
  1991).


\bibitem{KADANOFF_1}
{L.\,P.~Kadanoff}, {S.\,R.~Nagel}, L. Wu, and {S.\,M.~Zhou}, Phys.~Rev.~A {\bf
  39},  6524  (1989).

\bibitem{LUEB_4}
S.~L{\protect\"u}beck and {K.\,D.~Usadel}, Phys.~Rev.~E {\bf 56},  5138
  (1997).

\bibitem{KTITAREV_1}
{D.\,V.~Ktitarev} and {V.\,B.~Priezzhev}, Phys.~Rev.~E {\bf 58},  2883  (1998).



\bibitem{CHESSA_1}
A.~Chessa, E.~Marinari, A.~Vespignani, and S.~Zapperi, Phys.~Rev.~E {\bf 57},
  6241  (1998).

\bibitem{IVASH_unpub} E.\,V.~Ivashkevich, V.\,B.~Priezzhev, C.\,K.~Hu, and 
   C.\,Y.~Lin, to be published (1999).

\bibitem{DROSSEL_1} 
B.~Drossel, preprint cond-mat/9904075 v2 (1999).

\end{thebibliography}
\end{document}